\documentclass[
reprint,
twocolumn,
amsmath,amssymb,
aps,
pra
]{revtex4-2}

\usepackage{graphicx}
\usepackage{dcolumn}
\usepackage{bm}
\usepackage{color}
\usepackage{comment}
\usepackage{quantikz}
\usepackage{braket} 
\usepackage{amsmath}
\usepackage{amssymb}

\newcommand{\B}[1]{\textcolor{black}{#1}}
\newcommand{\Add}[1]{\textcolor{black}{#1}}
\newcommand{\Imp}[1]{\textcolor{black}{#1}}
\newcommand{\Blue}[1]{\textcolor{black}{#1}}
\newcommand{\Red}[1]{\textcolor{black}{#1}}
\newcommand{\Blu}[1]{\textcolor{black}{#1}}

\usetikzlibrary{backgrounds}

\begin{document}


\title{Fault-Tolerant Quantum Computing with High-Rate Symplectic Double Codes}

\author{Naoyuki Kanomata${}^{1,}$}
\email{naoyuki.kanomata@riken.jp}
\author{Hayato Goto${}^{1,2}$}

\affiliation{%
${}^1$RIKEN Center for Quantum Computing (RQC), Wako, Saitama 351-0198, Japan \\
${}^2$Corporate Laboratory, Toshiba Corporation, \Imp{Kawasaki, Kanagawa} 212-8582, Japan}%

\date{\today}

\begin{abstract}{
High-rate and large-distance quantum codes are expected to make fault-tolerant quantum computing more efficient, but most of them lack efficient fault-tolerant encoded-state preparation methods.
We propose such \B{fault-tolerant encoders for two symplectic double codes, $[[30,6,5]]$ and $[[36,12,5]]$. The advantage of
these codes is their compactness, namely, their small code block sizes for their high encoding rates and large distance}, allowing for early experimental
realization \B{of high-rate codes}.
Detecting crucial errors during encoding with as few auxiliary qubits as possible, our \B{encoders} can reduce resource overheads while keeping low logical error rates, compared to more naive
methods. Numerical simulations with a circuit-level noise model demonstrate the reliability and effectiveness of the proposed methods. We also develop \B{arbitrary-state encoders that enable} the
injection of arbitrary quantum states into the code \B{spaces}. Combined with basic fault-tolerant operations, \B{they support} universal quantum computation. We thus demonstrate that efficient and reliable
logical state preparation is achievable even for \B{compact and high-rate codes}, offering a potential
step toward efficient fault-tolerant quantum computing suitable for near-term experiments.
}\end{abstract}

\maketitle

\section{Introduction}
Quantum error correction (QEC) is a key requirement for realizing large-scale quantum computation.
Fault-tolerant quantum computing (FTQC)~\cite{1,2,3} applies QEC to protect quantum information from noises during computation.
So far, the surface code~\cite{4,5,6,7,8,9,10} has been the leading candidate for FTQC \Imp{because of its simple structure and high error threshold}.
However, one major limitation of the surface code is \Imp{its low encoding rate}, resulting in high resource overheads.

Recently high-encoding-rate quantum codes, such as quantum low-density parity-check (qLDPC) codes~\Imp{\cite{11}}, have drawn growing interest as promising alternatives.
These codes may allow us to reduce the overheads associated with FTQC by encoding more logical qubits in a code block.
Although these codes are typically more complex than the surface code, the recent progress in experimental systems such as trapped ions~\Add{\cite{Ryan2021a,Postler2022a,Moses2023a,Ryan2024a}} and neutral atoms~\Add{\cite{Graham2022a,Bluvstein2022a,Bluvstein2024a}} is enhancing the feasibility of their physical implementations.

\Imp{A major challenge for} qLDPC codes is the large block size requirement, which hinders near-term implementations.
For example, a typical small-size bivariate bicycle code, which is a leading candidate among qLDPC codes, has parameters $[[72,12,6]]$~\cite{28}, where $[[n,k,d]]$ means $k$ logical qubits encoded into $n$ physical qubits and the code distance of $d$.
Its modified version called the trivariate (multivariate) bicycle code can shrink the block size to $[[30,4,5]]$~\cite{19} while keeping the same error correction ability in terms of code distance. 
However, its encoding rate becomes lower from $12/72\approx 17\%$ to $4/30\approx 13\%$. 

In this work, we focus on \B{two Calderbank-Shor-Steane (CSS) codes~\cite{3,22,23}, $[[30,6,5]]$ and $[[36,12,5]]$,} derived from \B{$[[15,3,5]]$ and $[[18,6,5]]$ non-CSS stabilizer codes~\cite{18}, respectively,} via so-called symplectic double~\cite{16}.
\B{These symplectic double codes have higher encoding rates, ${6/30=20\%}$ and ${12/36\simeq 33\%}$, respectively,} than both the above bicycle codes while maintaining error-correction ability in terms of code distance.
Also note that previous studies on FTQC with symplectic double codes have focused on smaller distances up to $3$~\cite{20}.

Toward FTQC using the \B{high-rate symplectic double codes}, we propose \B{efficient fault-tolerant zero-state encoders, which are} based on the extension of Goto's minimum-overhead verification method for the Steane seven-qubit code~\cite{17} to \B{the [[30,6,5]] and [[36,12,5]] codes}.
We numerically evaluate the performance of the proposed \B{encoders} in a circuit-level noise model and show that \B{they} can achieve the notable reduction of necessary gates and ancilla qubits while keeping low logical error rates, compared to more naive methods.

Alongside designing the zero-state \B{encoders}, we also introduce \B{encoders} capable of embedding arbitrary quantum states into the code \B{spaces}.
\B{These arbitrary-state encoders enable} universal quantum computation, together with other basic operations through, e.g., so-called magic state distillation~\cite{26}.

Because of \B{their high encoding rates, small block sizes, and proposed encoders, these symplectic double codes} will be useful toward experimental demonstration of low-overhead FTQC in the near future.

\section{Results}
\subsection{Definition of the \B{symplectic double codes}}\label{II}
In general, symplectic double codes are defined as follows.
Consider the stabilizer matrix of a $[[n,k,d]]$ non-CSS code given by
\begin{align}
H = (H_{X} \mid H_{Z}),
\end{align}
where $H_{X}$ and $H_{Z}$ are binary matrices~\cite{3}.
\Imp{Since any two stabilizers must be commutative, the following symplectic orthogonality condition holds}:
\begin{align}
(H_X\ H_Z)(H_Z\ H_X)^T &= H_X H_Z^T + H_Z H_X^T 
 = O,
\end{align}
where $O$ denotes the zero matrix.
In this paper, binary vectors and matrices are defined over $GF(2)$\Imp{, that is, the addition with them is performed modulo 2}.
Using this property, we can construct a $[[2n, 2k, d]]$ CSS code by defining the matrices for the $X$ and $Z$ stabilizers, respectively, as
\begin{align}\label{(1)}
H'_{X} = (H_{X}\ H_{Z}), \quad H'_{Z} = (H_{Z}\ H_{X}).
\end{align}
Such a CSS code is referred to as a symplectic double code~\cite{16}.

\B{Hereafter we focus on the $[[30,6,5]]$ code for simplicity. All the corresponding results for the $[[36,12,5]]$ code are provided in the Appendix.}

\B{For the $[[30,6,5]]$ code}, we choose a $[[15,3,5]]$ non-CSS stabilizer code with the following stabilizer matrix~\cite{18}:
\begin{widetext}
\begin{align}
H= \left(
H_{X} \mid H_{Z} 
\right)= 
\left(
\begin{array}{ccccccccccccccc|ccccccccccccccc}
1 & 0 & 0 & 0 & 0 & 0 & 0 & 0 & 0 & 1 & 0 & 1 & 0 & 1 & 0 & 0 & 0 & 0 & 0 & 0 & 0 & 0 & 1 & 1 & 1 & 1 & 1 & 0 & 0 & 1 \\ 
0 & 0 & 0 & 0 & 0 & 0 & 0 & 1 & 1 & 1 & 1 & 1 & 0 & 0 & 1 & 1 & 0 & 0 & 0 & 0 & 0 & 0 & 1 & 1 & 0 & 1 & 0 & 0 & 1 & 1 \\ 
0 & 1 & 0 & 0 & 0 & 0 & 0 & 1 & 1 & 1 & 1 & 0 & 0 & 1 & 0 & 0 & 0 & 0 & 0 & 0 & 0 & 1 & 1 & 1 & 1 & 1 & 1 & 1 & 0 & 1 \\ 
0 & 0 & 0 & 0 & 0 & 0 & 1 & 1 & 1 & 1 & 1 & 1 & 1 & 0 & 1 & 0 & 1 & 0 & 0 & 0 & 0 & 1 & 0 & 0 & 0 & 0 & 1 & 1 & 1 & 1 \\ 
0 & 0 & 1 & 0 & 0 & 0 & 0 & 1 & 0 & 0 & 1 & 1 & 1 & 1 & 0 & 0 & 0 & 0 & 0 & 0 & 0 & 1 & 0 & 1 & 1 & 1 & 1 & 1 & 1 & 1 \\ 
0 & 0 & 0 & 0 & 0 & 0 & 1 & 0 & 1 & 1 & 1 & 1 & 1 & 1 & 1 & 0 & 0 & 1 & 0 & 0 & 0 & 1 & 1 & 1 & 1 & 0 & 0 & 0 & 0 & 1 \\ 
0 & 0 & 0 & 1 & 0 & 0 & 0 & 1 & 0 & 1 & 0 & 1 & 0 & 0 & 0 & 0 & 0 & 0 & 0 & 0 & 0 & 1 & 0 & 0 & 1 & 1 & 1 & 1 & 1 & 0 \\ 
0 & 0 & 0 & 0 & 0 & 0 & 1 & 0 & 0 & 1 & 1 & 1 & 1 & 1 & 0 & 0 & 0 & 0 & 1 & 0 & 0 & 1 & 1 & 0 & 0 & 1 & 0 & 1 & 1 & 0 \\ 
0 & 0 & 0 & 0 & 1 & 0 & 0 & 0 & 1 & 0 & 1 & 0 & 1 & 0 & 0 & 0 & 0 & 0 & 0 & 0 & 0 & 0 & 1 & 0 & 0 & 1 & 1 & 1 & 1 & 1 \\ 
0 & 0 & 0 & 0 & 0 & 0 & 0 & 1 & 0 & 0 & 1 & 1 & 1 & 1 & 1 & 0 & 0 & 0 & 0 & 1 & 0 & 0 & 1 & 1 & 0 & 0 & 1 & 0 & 1 & 1 \\ 
0 & 0 & 0 & 0 & 0 & 1 & 0 & 1 & 1 & 0 & 0 & 1 & 1 & 0 & 1 & 0 & 0 & 0 & 0 & 0 & 0 & 1 & 1 & 1 & 0 & 0 & 1 & 1 & 1 & 0 \\ 
0 & 0 & 0 & 0 & 0 & 0 & 1 & 1 & 1 & 0 & 0 & 1 & 1 & 1 & 0 & 0 & 0 & 0 & 0 & 0 & 1 & 1 & 0 & 0 & 0 & 0 & 0 & 0 & 1 & 1    
\end{array}
\right).
\end{align}
\end{widetext}
We can verify that the commutativity condition holds:
\begin{align}
H'_X H^{\prime T}_Z = 
(H_{X}\ H_{Z})(H_{Z}\ H_{X})^T = O_{12,12},
\end{align}
where $O_{i,j}$ denotes the $i\times j$ zero matrix.
Thus we obtain a $[[30,6,5]]$ symplectic double code.
\Blue{One potential drawback of this code is the relatively high weights of its stabilizers, up to 14.
This does not fundamentally limit the code’s fault-tolerant properties by using Steane- or Knill-style QEC,
but it increases the overheads for syndrome extraction, namely, stabilizer measurements. 
We should take the overheads into account when implementing the present scheme. }

The logical Pauli operators for the $[[30,6,5]]$ code are derived as follows.
Let $M_X$ and $M_Z$ denote the binary matrices corresponding to the logical $X$ and $Z$ operators, respectively.
These matrices must satisfy the following commutation and orthogonality conditions:
\begin{align}
H'_X M_Z^T = O_{12,6}, \quad H'_Z M_X^T = O_{12,6}, \quad M_X M_Z^T = I_{6},
\end{align}
where $I_i$ denotes the $i\times i$ identity matrix.
The first two conditions ensure that the logical operators commute with all the stabilizers.
The final condition ensures the correct \Imp{commutation} and anti-commutation relations of \Imp{the} logical $X$ and $Z$ operators.
\Imp{By finding three vectors orthogonal to and linearly independent of both $H_{X}$ and $H_{Z}$, we obtain the explicit forms of $M_X$ and $M_Z$ as}
\begin{align}
M_{X} = M_{Z} =
\begin{pmatrix} M & O_{3,15} \\ O_{3,15} & M \end{pmatrix},
\label{eq-MxMz}
\end{align}
where 
\begin{align}
M=\left(\begin{array}{ccccccccccccccc}
1&0&0&1&0&1&1&0&0&1&1&0&1&0&0\\
0&1&0&0&0&1&0&1&0&1&0&0&0&1&0\\
0&0&1&0&1&1&0&0&1&1&0&1&0&0&1 
\end{array}
\right)
\label{eq-M}
\end{align}
consists of \Imp{the above three vectors}.
The logical Pauli operators are expressed more explicitly as follows:
\begin{align}
Z_{L1} =& Z_{1}Z_{4}Z_{6}Z_{7}Z_{10}Z_{11}Z_{13}\Imp{,}\\
X_{L1} =& X_{1}X_{4}X_{6}X_{7}X_{10}X_{11}X_{13}\Imp{,}\\
Z_{L2} =& Z_{2}Z_{6}Z_{8}Z_{10}Z_{14}\Imp{,}\\
X_{L2} =& X_{2}X_{6}X_{8}X_{10}X_{14}\Imp{,}\\
Z_{L3} =& Z_{3}Z_{5}Z_{6}Z_{9}Z_{10}Z_{12}Z_{15}\Imp{,}\\
X_{L3} =& X_{3}X_{5}X_{6}X_{9}X_{10}X_{12}X_{15}\Imp{,}\\
Z_{L4} =& Z_{16}Z_{19}Z_{21}Z_{22}Z_{25}Z_{26}Z_{28}\Imp{,}\\
X_{L4} =& X_{16}X_{19}X_{21}X_{22}X_{25}X_{26}X_{28}\Imp{,}\\
Z_{L5} =& Z_{17}Z_{21}Z_{23}Z_{25}Z_{29}\Imp{,}\\
X_{L5} =& X_{17}X_{21}X_{23}X_{25}X_{29}\Imp{,}\\
Z_{L6} =& Z_{18}Z_{20}Z_{21}Z_{24}Z_{25}Z_{27}Z_{30}\Imp{,}\\
X_{L6} =& X_{18}X_{20}X_{21}X_{24}X_{25}X_{27}X_{30},
\end{align}
where $X_{Li}$ and $Z_{Li}$ are the Pauli operators for the $i$-th logical qubit, and $X_{j}$ and $Z_{j}$ are those for the $j$-th physical qubit.

For CSS codes, Clifford operations can often be performed 
\Imp{fault-tolerantly by so-called transversal gates~\cite{3}}.
For example, transversal logical CNOT gates can be performed \Imp{by} transversal physical CNOT gates for arbitrary CSS codes.

\begin{figure}[tbp]
\includegraphics[width=\columnwidth]{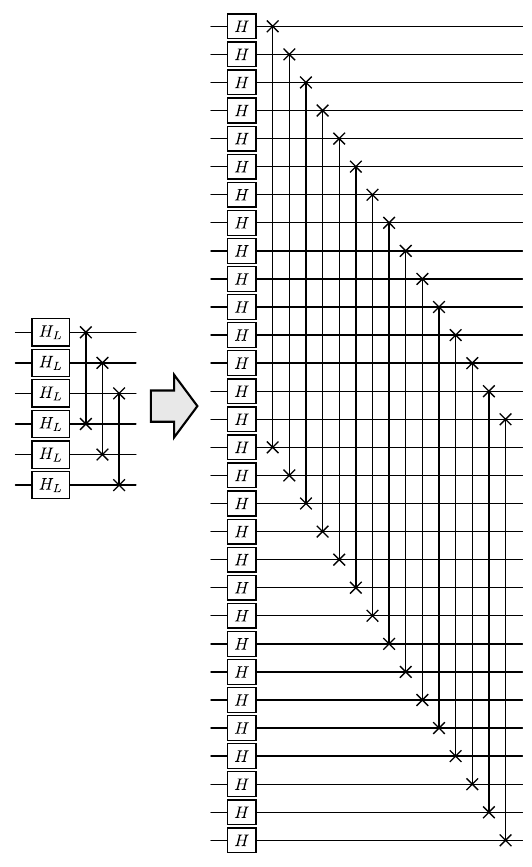}
\caption{\textbf{Logical Hadamard gates followed by logical SWAP gates for the $[[30,6,5]]$ symplectic double code.}}
\label{fig6}
\end{figure}

Next, we consider logical Hadamard gates.
By the definitions of $H'_X$ and $H'_Z$
\Blue{in Eq.~\Blu{(\ref{(1)})}}, the transversal physical Hadamard gate followed by the \Imp{physical SWAP gates} between the \Imp{$q$-th} and $(q+15)$-th physical qubits $(q=1,\ldots,15)$ exchanges the $X$ and $Z$ stabilizers, 
\Blue{resulting in some logical gates, 
which are known as 
Hadamard-type fold-transversal gates 
based on the $ZX$-duality of this code~\cite{20,21}}.
Under the same transformation, the logical $X$ and $Z$ operators are also exchanged, accompanied by the \Blue{logical SWAP gates} 
between the \Imp{$Q$-th} and $(Q+3)$-th logical qubits $(Q=1, 2, 3)$, from the definitions of $M_{X}$ and $M_{Z}$ 
\Blue{in Eq.~(7)}.
\Blue{In summary,} 
the transversal physical \Imp{Hadamard} gate followed by the physical SWAP \Imp{gates} is equivalent to 
the transversal logical Hadamard gate followed by the logical \Imp{SWAP gates}, as shown in Fig.\ref{fig6}.
\subsection{Zero-state and Plus-state encoders}\label{III}
\begin{figure}[tbp]
\includegraphics[width=\columnwidth]{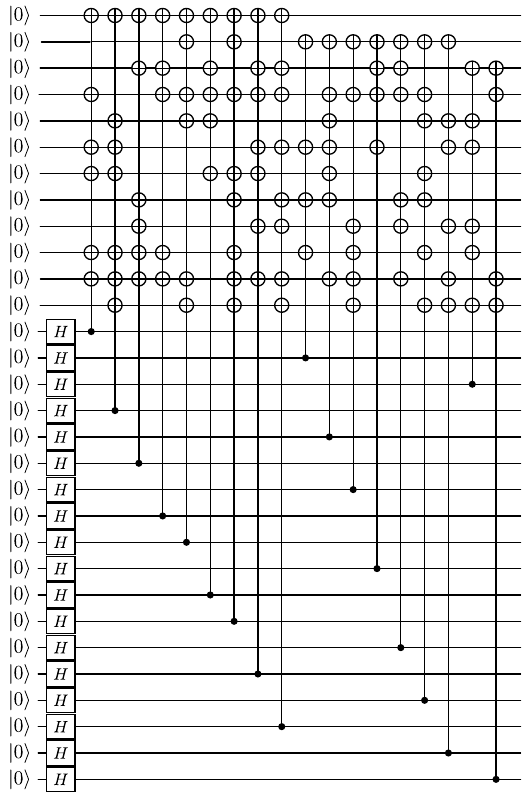}
\caption{\textbf{Plus-state encoder for the $[[30,6,5]]$ symplectic double code with 108 CNOT gates.}}
\label{fig1}
\end{figure}

For the reason mentioned later, 
here we first focus on the plus-state encoder of 
the $[[30,6,5]]$ code, 
where the plus state is the eigenstate of $X$, 
namely, $|+\rangle=(|0\rangle + |1\rangle)/\sqrt{2}$.
Using the matrix $H'_Z$, we can construct the plus-state encoder
as follows.
By the Gaussian elimination, 
we first convert $H'_Z$ to the form such that 
its left-most part is the identity matrix $I_{12}$.
Correspondingly, 
we prepare the first 12 physical qubits 
in the eigenstate of $Z$, namely, $|0\rangle$, 
and the other 18 qubits in $|+\rangle$ by Hadamard gates.
Then, we perform physical CNOT gates 
each of which corresponds to a row vector of $H'_Z$. 
That is, if the row vector has 1 on the $i$th and $j$th qubits ($i\le 12$ and $j>12$), 
we perform the corresponding CNOT gate 
whose target and control qubits are 
the $i$th and $j$th qubits, respectively.
Thus, the 30-qubit state finally becomes the stabilizer state\Imp{, namely, the simultaneous eigenstate with eigenvalues of $+1$, with respect to} not only all the $Z$ stabilizers, 
but also all the $X$ stabilizers and 
all the logical $X$ operators, 
which is the \Imp{logical} all-plus state of the $[[30,6,5]]$ code.
The resultant plus-state encoder is shown in Fig.~\ref{fig1}.
\begin{figure*}[tbp]
\includegraphics[width=2\columnwidth]{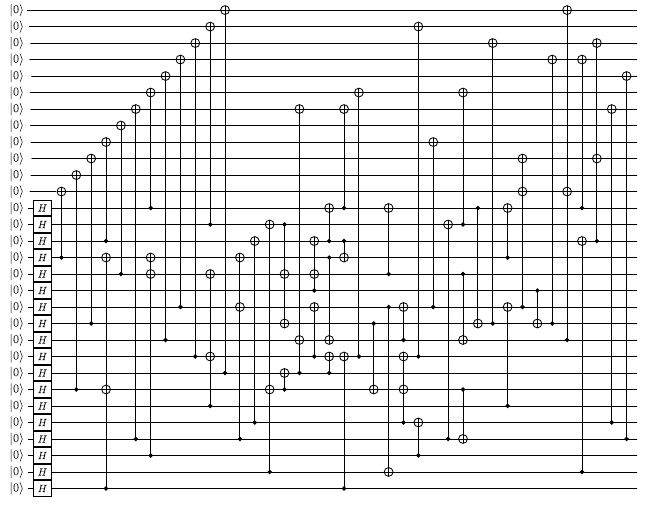}
\caption{\textbf{Plus-state encoder for the $[[30,6,5]]$ symplectic double code with 67 CNOT gates.}}
\label{fig2}
\end{figure*}
The above encoder needs 108 CNOT gates.
To reduce them, we next apply the overlap method 
proposed in~\cite{27} to the encoder.
The point of the overlap method is as follows. 
If two column vectors, e.g., 
the $j$th and $l$th column vectors of $H'_Z$ 
(\Imp{${j, l > 12}$}) have two or more overlaps in terms of 1, 
then there is a possibility for reducing the number of CNOT gates by performing a CNOT gate on \Imp{the $j$th and $l$th qubits}, 
instead of the above CNOT gates on them.
Optimizing such CNOT gates, 
we succeeded in reducing the number of CNOT gates 
from $108$ to $67$, as shown in Fig.~\ref{fig2}.

In a similar manner to the plus-state encoder 
in Fig.~\ref{fig2}, 
we can construct the zero-state encoder using $H'_X$ \Imp{and the overlap method}.
\Imp{However,}   
the logical all-zero state obtained by \Imp{such an} encoder 
\Imp{exhibited} lower performance than 
the logical all-plus state obtained 
by the encoder in Fig.~\ref{fig2} 
under evaluation with a circuit-level noise model 
used \Red{in the “Simulation of logical CNOT gates” section}.
(The performance difference may come from \B{our specific choice of the qubits for the Gaussian elimination such that the left-most part of a check matrix is converted into the identity matrix}.)
This is the reason why we started with
the plus-state encoder.
Instead of using $H'_X$, 
we can prepare the all-zero state 
using the plus-state encoder followed 
by the logical Hadamard and SWAP gates shown in Fig.~\ref{fig6}, 
resulting in another zero-state encoder.
We use it as a zero-state encoder throughout this work.

\subsection{Decoder}
In this work, we developed a decoder based on a lookup table \Imp{specifically} designed for the $[[30,6,5]]$ symplectic double code.
This is achievable because of its compact structure, where the ``compactness" means that the number of syndromes used for decoding is only $12$, resulting in only $2^{12}$ rows of the lookup table.
\Blue{(It has turnded out that 
we must consider error patterns up to 
weight 4 to complete the $2^{12}$ rows.)}
This decoder is optimal in the sense that the syndrome information is fully utilized.

The lookup table for the logical $X$ measurement is built as follows.
First, we generate all possible phase-flip error patterns up to weight \Imp{4}, and compute the corresponding syndromes as follows:
\begin{align}
\mathbf{s}_{j} = H'_{X}\mathbf{e}_{j},
\end{align}
where $\mathbf{e}_{j}$ and $\mathbf{s}_{j}$ are the binary vector and syndrome values, respectively, corresponding to the $j$th phase-flip error pattern.
\Imp{If different error patterns give the same syndrome value, we choose the lowest-weight one. 
Thus, every syndrome value and its associated error pattern} are cataloged, resulting in the lookup table for the logical $X$ measurement.
The lookup table for the logical $Z$ measurement can \Imp{also} be constructed with $H'_{Z}$, instead of $H'_{X}$, in a similar manner.

The table provides the most likely error pattern for \Imp{a} syndrome result.
The decoder utilizes the table to find the most likely error pattern, where the syndrome result is used as the index of the error pattern.

We evaluate the performance of our decoder by numerical Monte-Carlo simulations, 
\Blue{where we assume the code capacity noise model, 
that is, there are only physical-qubit memory errors 
and no other kinds of errors.}
(\Blue{Such code capacity experiments can be done 
by generating an error vector and decoding it with 
the code's parity-check matrix. 
However, we use stabilizer simulations to verify our proposed encoding circuit. 
In this work, we used Stim~\cite{15} for stabilizer simulations.})
We first create \Blue{an} error-free logical all-\Imp{plus} state of the symplectic double code, utilizing the encoder in Fig.~\ref{fig2}.
Then, independent phase-flip errors are applied to physical qubits, with \Imp{probability} of $p_{\mathrm{flip}}$.
Following that, we carry out ideal measurements on all the physical qubits in the $X$ basis.
We decode these outcomes using the lookup table explained above.
We consider that the decoding has succeeded if \Imp{all the decoding outcomes for the six logical qubits are $|+\rangle$} and otherwise it has failed.
\Blue{The decoding error probability was estimated through 
112,000,000 trials 
for $p_{\mathrm{flip}}=0.001$ 
and 11,200,000 trials otherwise}.
\begin{figure}[tbp]
\includegraphics[width=\columnwidth]{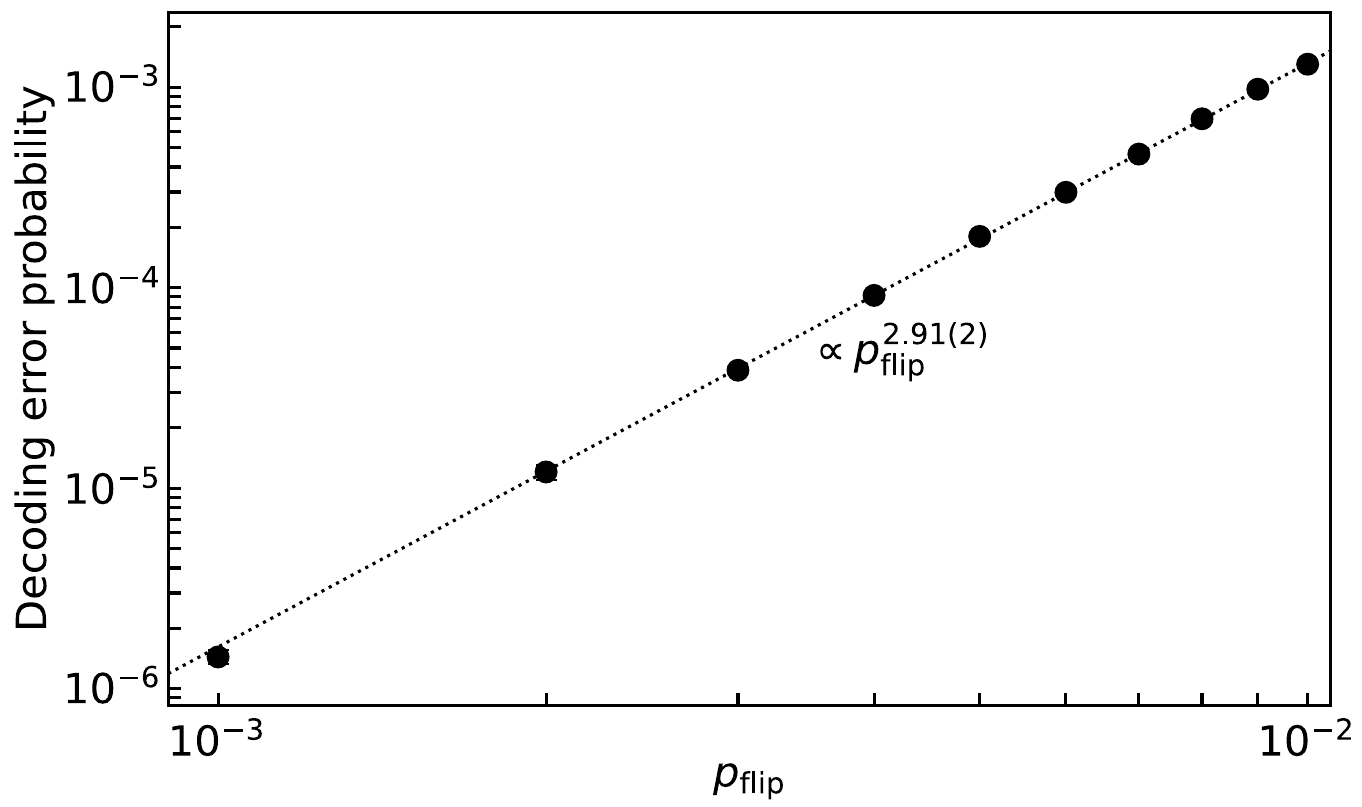}
\caption{\textbf{Simulation results of the decoding error probability 
\B{for the $[[30,6,5]]$ symplectic double code} 
under phase-flip $(Z)$ errors with probability of $p_{\mathrm{flip}}$.}}
\label{fig3}
\end{figure}
Figure~\ref{fig3} shows the results of the decoding \Imp{error} probability as a function of $p_{\mathrm{flip}}$.
The \Imp{decoding error probability} follows a \Imp{power law} proportional to 
$p_{\text{flip}}^{\Blu{2.91}}$. The exponent \Blu{2.91} is in reasonable agreement with the theoretically expected value based on the code distance $5$, namely, $(5+1)/2=3$, indicating that the decoder is optimal, as expected.
\Blue{(A distance-5 code can correct all weight-2 errors and 
cannot correct some weight-3 errors, 
resulting in ${\Theta (p_{\text{flip}}^3)}$ 
logical error probability.)}

\subsection{Fault-tolerant zero-state and plus-state encoders}
\label{sec-FTencoder}
Note that the encoders presented \Red{in the “Decoder” section} are not fault-tolerant \Imp{because of correlated multi-qubit errors due to CNOT gates}.
Here we make the encoder in Fig.~3 fault-tolerant by generalizing Goto's minimum-overhead verification method for the Steane code~\cite{17}.
This method achieves fault tolerance by first preparing a logical state using a non-fault-tolerant encoder and then eliminating potentially harmful errors through appropriate logical Pauli measurements.
Here we focus on a fault-tolerant encoder for the \Imp{all-plus} state.
As explained \Red{in the “Decoder” section}, the logical \Imp{all-zero} state can be prepared by transversal Hadamard gates on the \Imp{all-plus} state, which is fault-tolerant.
\B{
We first rearrange the physical CNOT gates in Fig.~\ref{fig2} 
so that the instantaneous state at each step is a stabilizer state with as low stabilizer weights as possible. 
Next, we find a special step at which all the stabilizer weights 
are less than or equal to 3, but some of them exceed 3 
at the next step.
Such low-weight stabilizers 
allow us to regard the correlated errors as single-qubit errors~\cite{17,Goto2023a}, 
and therefore we care about only the correlated errors after 
the special step, 
similarly to ``hook" errors in the surface-code syndrome measurements, where only correlated errors due to the third and fourth CNOT gates are harmful~\cite{5}.}

Next, \Blue{we search for logical-$X$ measurements 
that can catch all the correlated errors caused by 
CNOT gates after the above special step, 
where we minimize the number of the logical-$X$ measurements 
and the weights of the logical $X$ operators. 
This search results} in the verification circuit in Fig.~5A. 
We accept the output state if all the measurement outcomes are zero.
\Add{
While the above verification can eliminate harmful $Z$ errors, we also have to eliminate harmful $X$ errors. To achieve that, we use the circuit in Fig~5B, where we first prepare two logical all-plus states using the circuit in Fig.~5A, then perform transversal logical CNOT gates and measure the second logical block in the $Z$ basis. 
We accept the logical state in the first block if no errors are detected, that is, if all the stabilizer and logical values of the measurement outcomes are zero.
Our proposed fault-tolerant plus-state encoder thus consists of the circuits in Figs.~5A and 5B.
}
\begin{figure}[htbp]
\includegraphics[width=0.94\columnwidth]{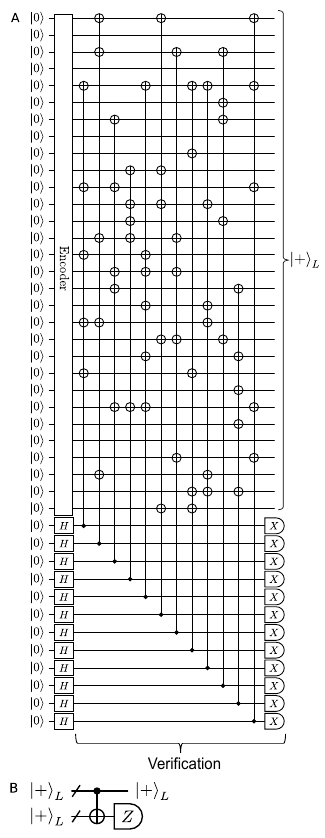}
\caption{\textbf{Fault-tolerant plus-state encoder for the $[[30,6,5]]$ symplectic double code.} (\textbf{A}) Plus-state encoder utilizing \Imp{Goto's verification method~\cite{17} to eliminate harmful $Z$ errors}. 
The ``Encoder" box is the \Imp{quantum circuit} shown in Fig.~\ref{fig2}. 
(\textbf{B}) \Imp{Verification procedure to eliminate harmful $X$ errors. The two input states are prepared by the encoder in (A).}}
\label{fig4}
\end{figure}
\Add{
For comparison, we also evaluate more naive fault-tolerant encoders shown in Figs.~6A--6C, where many copies of all-plus states are prepared with the encoder in Fig.~\ref{fig2} and harmful $Z$ and $X$ errors are eliminated by verification with the copies~\cite{25}.
}
\begin{figure}[htbp]
\includegraphics[width=0.68\columnwidth]{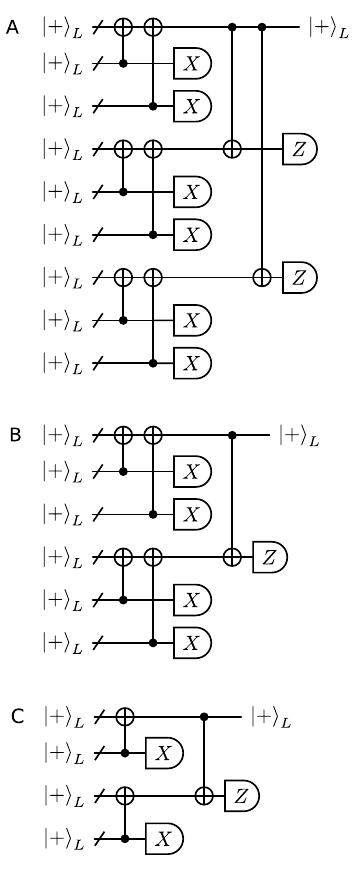}
\caption{\textbf{Naive fault-tolerant plus-state encoders. }
(\textbf{A}) 9-copy encoder. We use three copies for the verification to eliminate both $Z$ and $X$ errors. The input copies are prepared by the encoder in Fig.~\ref{fig2}. 
We accept the output state if no error is detected, otherwise restart the encoding. 
(\textbf{B}) 6-copy encoder. We use three copies for the verification to eliminate $Z$ errors and two copies for $X$ errors. 
(\textbf{C}) 4-copy encoder. We use two copies for the verification to eliminate both $Z$ and $X$ errors.}
\label{fig7}
\end{figure}
\subsection{Simulation of logical CNOT gates}
\label{sec-logicalCNOT}
\Imp{Here} we evaluate the performance of the encoders presented \Red{in the “Fault-tolerant zero-state and plus-state encoders” section} through numerical simulations of \Imp{logical CNOT gates}. 
\Imp{The simulation method is depicted in Fig.~7A~\cite{24}, where faulty transversal logical CNOT gates followed by QEC gadgets~\cite{12,13,14} are performed on two ideal logical Bell states. In this work, we perform QEC by error-correcting teleportation depicted in Fig.~7B. After disentangling the Bell states and measuring all the qubits, the outcomes are decoded with the lookup tables.}
\begin{figure*}[htbp]
\includegraphics[width=2\columnwidth]{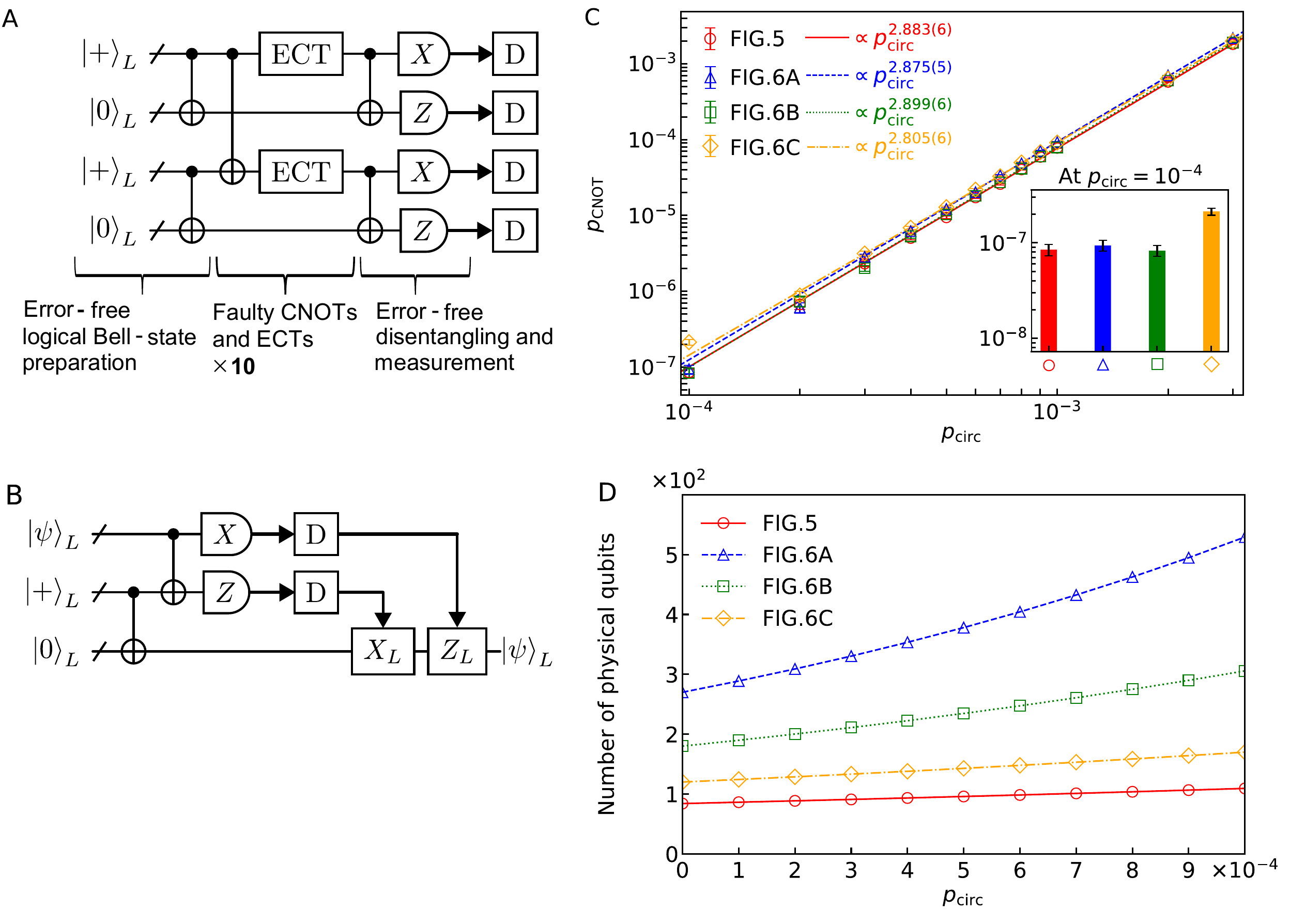}
\caption{
\textbf{Evaluation of the encoders and logical CNOT gates for the $[[30,6,5]]$ symplectic double code.}
(\textbf{A}) Simulation method for estimating logical-CNOT error probability $p_{\text{CNOT}}$~\cite{24}.
We first create two error-free logical Bell states encoded with the $[[30,6,5]]$ symplectic double code. On the Bell states, we perform ten sets of transversal logical CNOT gates followed by QEC gadgets, which are implemented by faulty transversal physical CNOT gates and error-correcting teleportation (ECT)~\cite{12,13,14}, respectively.
Following this, the logical states are disentangled with error-free CNOT gates.
Finally, all physical qubits are ideally measured in the $Z$ and $X$ bases. 
Measurement outcomes are decoded with the lookup tables.
If all decoded logical outputs are consistent with the input states, we consider that the ten transversal logical CNOT gates have succeeded and otherwise they have failed, the failure probability of which is denoted by $p_{10}$.
The logical CNOT error probability is evaluated as $p_{\mathrm{CNOT}} = 1 - (1 - p_1)^{1/6}$ with $p_1 = 1 - (1 - p_{10})^{1/10}$.
(\textbf{B}) Error-correcting teleportation (ECT)~\cite{12,13,14} used in (A). 
The logical Bell states \Blue{for ECT} are prepared with one of the encoders in Figs.~5 and 6 in the circuit-level noise model described in the main text, 
\Blue{where the preparation of 
two logical all-plus states 
used for the logical Bells state 
is repeated until success\B{, namely, passing the verification stage with no detected errors}. 
(Memory errors during waiting time until success 
are not taken into account in this work.)}
The CNOT gates and measurements for the logical Bell measurement are also faulty.
(\textbf{C}) Numerical results of the logical-CNOT error probability as a function of the error rate, $p_{\mathrm{circ}}$, of the noise model.
(\textbf{D}) \Blue{Average} total number of physical qubits required for preparing a logical all-plus state 
\B{passing the verification stage with no detected errors}.
}
\label{fig10}
\end{figure*}

\Add{
In this work, we apply the following circuit-level noise model~\cite{24} to the faulty logical CNOT gates and QEC gadgets.
The error rate is denoted by $p_{\mathrm{circ}}$.
Each physical CNOT gate is implemented by an ideal CNOT gate followed by one of 15 non-identity two-qubit Pauli errors with probability of $p_{\mathrm{circ}}/15$. 
Each physical-qubit preparation in $|0\rangle$ is followed by a bit-flip error with probability of $p_{\mathrm{circ}}$.
Before each physical-qubit measurement in the $Z$ basis, a bit-flip error occurs with probability of $p_{\mathrm{circ}}$. 
We assume no memory and single-qubit-gate errors. This error model may be relevant for some physical systems such as trapped ions~\Add{\cite{Ryan2021a,Postler2022a,Moses2023a,Ryan2024a}} and neutral atoms~\Add{\cite{Graham2022a,Bluvstein2022a,Bluvstein2024a}}.
}

Figure~\ref{fig10}C shows that the logical-CNOT error probability $p_{\text{CNOT}}$, 
\Blue{which was estimated through 
11,200,000 trials 
for $p_{\mathrm{circ}}=0.0001$ 
and 1,120,000 trials otherwise}, 
follows a power law as $p_{\text{CNOT}} \propto p_{\text{circ}}^{a}$. 
For all the encoders, 
the estimated values of the exponent $a$ are close to 3, 
which is consistent with the code distance of 5. 
\B{This demonstrates the fault tolerance of the encoders.}
Moreover, the proposed encoder achieved the lowest $p_{\text{CNOT}}$ at $p_{\text{circ}}=10^{-4}$, 
as shown in the inset in Fig.~\ref{fig10}C.
\B{The logical-CNOT error probability of $10^{-7}$ at $p_{\text{circ}}=10^{-4}$ is much lower than 
that for the distance-3 Steane code, 
which is estimated at $10^{-5}$~\cite{17}. 
This shows the advantage of the large-distance symplectic double codes.
(The logical-CNOT error probabilities for the $[[36,12,5]]$ code are comparable to those for the $[[30,6,5]]$ code. 
See the Appendix for details.)}

Additionally, we numerically estimated the \Blue{average} total number of physical qubits required for the preparation of a logical all-plus state \B{passing the verification stage with no detected errors}, 
where the repetition due to postselection 
is taken into account. 
The results are shown in Fig.~\ref{fig10}D, 
which shows that the proposed encoder in Fig.~5 can achieve \Blue{not only} remarkably fewer physical qubits 
but also a smaller slope, namely, 
higher acceptance probability than the naive encoders in Figs.~6A--6C, as expected. 
\B{The small slope indicates the small impact of the post-selection in the proposed encoder.}

\vspace{1\baselineskip}

\subsection{Arbitrary-state encoder}
To achieve universal fault-tolerant quantum computation, e.g., via magic-state distillation~\cite{26},
we need an arbitrary-state encoder 
to inject a magic state into the code space.
For this purpose, 
here we propose an arbitrary-state encoder 
for the $[[30,6,5]]$ symplectic double code.
\Blue{(Note that this encoder is not fault-tolerant. 
Therefore, to achieve fault tolerance, 
we need appropriate verification like magic state distillation 
after logical-state preparation using this encoder.)}

\begin{figure*}[htbp]
\includegraphics[width=1.9\columnwidth]{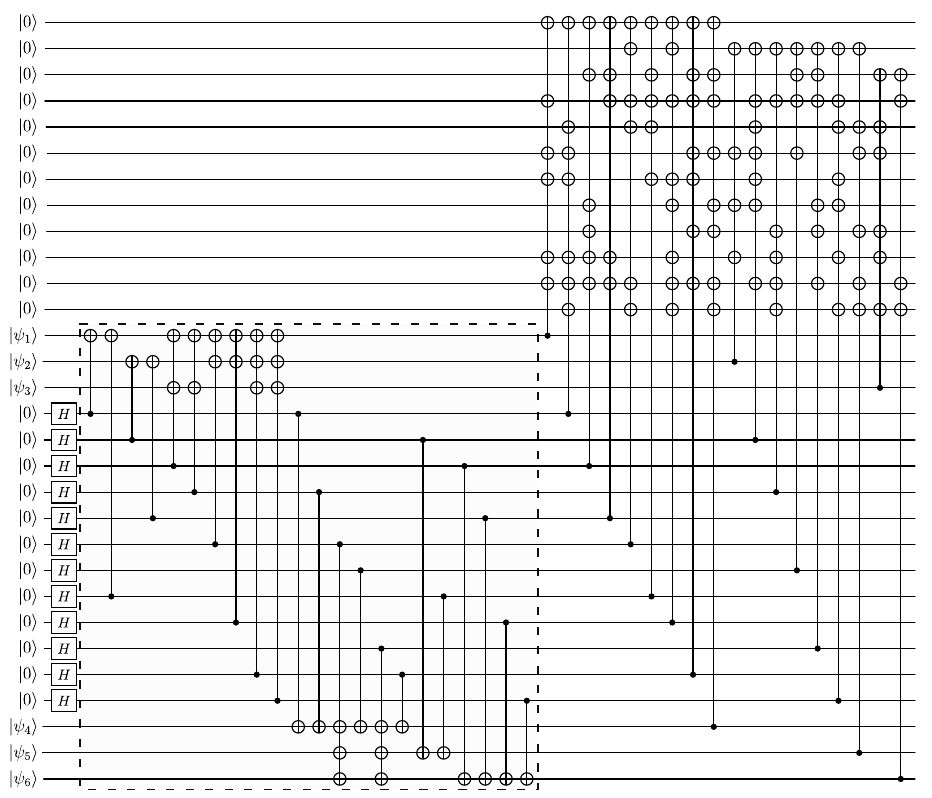}
\caption{\textbf{Arbitrary-state encoder for the $[[30,6,5]]$ symplectic double code.}}
\label{fig14}
\end{figure*}

Figure~\ref{fig14} shows the proposed encoder, 
where $|\psi_j \rangle$ (${j=1, 2, \ldots , 6}$) are 
six input states to be converted into 
the corresponding logical states $|\psi_j \rangle_L$.
(For simplicity, this figure shows the case of a product state, 
but this encoder can convert arbitrary input states including entangled states.)
This encoder is obtained by adding appropriate CNOT gates enclosed by the dashed rectangle 
between the Hadamard and CNOT gates in 
the plus-state encoder in Fig.~\ref{fig1}.
We determined the added CNOT gates using 
the following matrix $M'_Z$ obtained 
from $M_Z$ in Eq.~(7) by the Gaussian elimination 
with $H'_Z$:
\begin{align}
M'_Z
&=
\begin{pmatrix}
O_{3,12} & M' \\
O_{3,15} & M
\end{pmatrix},
\\
M'
&=
\left(
\begin{array}{cccccccccccccccccc}
1&0&0&1&0&1&1&0&1&0&1&1&0&1&1&0&0&0\\
0&1&0&0&1&0&0&1&1&0&0&1&0&1&1&0&0&0\\
0&0&1&0&0&1&1&0&0&0&0&0&0&1&1&0&0&0        
\end{array}
\right),
\end{align}
where $M$ is the matrix in Eq.~\Blu{(\ref{eq-M})}.
By its definition, $M'_Z$ defines logical $Z$ operators. 
\B{Each added CNOT gate corresponds to a pair of 1's in a row of $M'_Z$, 
where one of the pair (the target qubit) corresponds to one of the input qubits 
(the 13th, 14th, 15th, 28th, 29th, or 30th column)
and the other (the control qubit) corresponds to one of the other qubits 
(one of the 16th to 27th columns).}
In the following, we explain the validity of this encoder.

First, when all the input states are $|+\rangle$,
the added CNOT gates does not affect the 30-qubit state, 
and consequently this circuit is equivalent to 
that in Fig.~\ref{fig1}.
Thus, we successfully obtain the logical all-plus state 
$|+\rangle_L^{\otimes 6}$.

Next, consider the case where all the input states are $|-\rangle = Z|+\rangle$.
Then, we successfully obtain the logical all-minus state
$|-\rangle_L^{\otimes 6}$ as follows:
\begin{align}
&
U_{\mathrm{CNOT}}
U'_{\mathrm{CNOT}}
|0\rangle^{\otimes 12}
|-\rangle^{\otimes 3}
|+\rangle^{\otimes 12}
|-\rangle^{\otimes 3}
\nonumber \\
=&
U_{\mathrm{CNOT}}
U'_{\mathrm{CNOT}}
Z_{13} Z_{14} Z_{15} Z_{28} Z_{29} Z_{30} 
|0\rangle^{\otimes 12}
|+\rangle^{\otimes 18}
\nonumber \\
=&
U_{\mathrm{CNOT}}
Z_L^{\otimes 6}
U'_{\mathrm{CNOT}}
|0\rangle^{\otimes 12}
|+\rangle^{\otimes 18}
\nonumber \\
=&
Z_L^{\otimes 6}
U_{\mathrm{CNOT}}
|0\rangle^{\otimes 12}
|+\rangle^{\otimes 18}
\nonumber \\
=&
Z_L^{\otimes 6}
|+\rangle_L^{\otimes 6}
\nonumber \\
=&
|-\rangle_L^{\otimes 6},
\label{eq-minus}
\end{align}
where $U_{\mathrm{CNOT}}$ and $U'_{\mathrm{CNOT}}$ 
are the unitary operators corresponding to the CNOT gates in Fig.~\ref{fig1} 
and the added CNOT gates, respectively, and 
$Z_L^{\otimes 6}$ are the six logical $Z$ operators defined \Imp{by} $M'_Z$.
In Eq.~\Blu{(\ref{eq-minus})}, 
we have used the following 
\Imp{three facts. First, 
by definition of $M'_Z$, 
$U'_{\mathrm{CNOT}}
Z_{13} Z_{14} Z_{15} Z_{28} Z_{29} Z_{30} 
=
Z_L^{\otimes 6}
U'_{\mathrm{CNOT}}$. 
Second, 
since $U'_{\mathrm{CNOT}}$ 
does not affect its target qubits in $|+\rangle$, 
$U'_{\mathrm{CNOT}}
|0\rangle^{\otimes 12}
|+\rangle^{\otimes 18}
=
|0\rangle^{\otimes 12}
|+\rangle^{\otimes 18}$. 
Third, 
since $U_{\mathrm{CNOT}}$ commutes with $Z_L^{\otimes 6}$, 
$U_{\mathrm{CNOT}} Z_L^{\otimes 6}
=
Z_L^{\otimes 6}U_{\mathrm{CNOT}}$.}
Similarly, \Imp{any product state of $|+\rangle$ and $|-\rangle$ 
is} converted into their corresponding logical states by this encoder.

Finally, from the linearity of unitary operators, 
this encoder converts any superposition of $|\pm \rangle$ product states 
into its corresponding logical state.
Therefore, the quantum circuit in Fig.~\ref{fig14} 
is the \Imp{desired} arbitrary-state encoder for the $[[30,6,5]]$ symplectic double code.

\B{This arbitrary-state encoder is not fault-tolerant. 
Making it fault-tolerant and evaluating the resulting encoder, e.g., for magic states in a circuit-level noise model are left for future work.}

\section{Discussion}
In this work, we have proposed \B{efficient fault-tolerant encoders for two compact and high-rate symplectic double codes,  \([[30,6,5]]\) and  \([[36,12,5]]\)}.
Leveraging \B{these encoders}, we have demonstrated low-overhead and low-error-rate logical CNOT gates under a circuit-level noise model.
To enable universal quantum computation with the \B{symplectic double codes}, we have also developed \B{arbitrary-state encoders for these codes}.
Further optimizing the encoding circuits for \B{these codes} remains an open challenge.
The arbitrary-state \B{encoders} together with 
transversal CNOT and Hadamard gates 
\B{enable} transversal logical phase gates $S$ and 
non-Clifford gates 
via state distillation and gate teleportation~\cite{24,44}.
(More efficient magic state preparation beyond distillation, 
e.g., using direct measurements of logical operators, 
which has been realized for the Steane code~\cite{17}, 
is desired but left as a future challenge.)
To achieve addressable logical gates beyond transversal ones, 
we need fault-tolerant preparation of 
appropriate logical Clifford states 
\B{such as partial $|+\rangle$ states~\cite{24}}, 
which is possible by the arbitrary-state \B{encoders} 
followed by verification like Fig.~\ref{fig7}A.
Toward more efficient logical computation, 
logical gates originating from the $ZX$-duality of 
the symplectic double code~\cite{20}, 
such as a fold-transversal CZ gate~\cite{21} 
(transversal application of CZ gates 
between the first and second halves of the code, 
resulting in a transversal logical CZ gate 
between the first three and second three logical qubits), 
will be helpful.

Looking ahead, a critical next step is the experimental implementation of the proposed protocols on physical quantum hardware platforms, such as trapped-ion and neutral-atom systems.
The proposed encoders will be helpful for near-term experimental demonstration of low-overhead FTQC with the \B{symplectic double codes}.
Such experiments will also allow us to evaluate the performance of the proposed encoders under realistic noises. 

Another interesting direction is scaling up the code size beyond the [[30,6,5]] \B{and [[36,12,5]] codes}, 
which \B{were} chosen because of 
the suitable \B{sizes} for 
fault tolerance analysis and 
early experimental realization. 
Recently, AI-driven discovery of efficient encoding circuits has been reported~\cite{PRX}. 
However, 
the codes have been limited to single-logical-qubit ones 
or high-rate but small-size ones. 
Toward scaling up, 
the combination of such an AI-driven approach 
with the present methods is promising. 

\section*{Acknowledgments}
N.K. would like to express his gratitude to Ryota Nakai for providing valuable advices in the course of conducting this work.
This work was supported by JST Moonshot R\&D, Grant No. JPMJMS256E.
Parts of numerical simulations were done on the HOKUSAI supercomputer at RIKEN (project ID RB230022).

\appendix
\section{Results for the $[[36,12,5]]$ Symplectic Double Code}
In the main text, we present the results for the $[[30,6,5]]$ symplectic double code.
To \B{show that the proposed methods are not limited to this specific code}, we provide the corresponding results for the $[[36,12,5]]$ symplectic double code in the Appendix.

The $[[36,12,5]]$ symplectic double code is obtained from \B{an} $[[18,6,5]]$ non-CSS stabilizer code with the following stabilizer matrix~\cite{18}:
\begin{widetext}
\begin{align}
H= \left(
H_{X} \mid H_{Z} 
\right)= 
\left(
\begin{array}{cccccccccccccccccc|cccccccccccccccccc}
1&0&0&0&0&1&0&1&1&0&0&1&1&0&1&0&0&1&0&0&0&0&0&0&0&0&1&0&0&0&0&1&0&1&0&0\\
0&0&0&0&0&0&0&0&1&0&0&0&0&1&0&1&0&0&1&0&0&0&0&1&0&1&0&0&0&1&1&1&1&1&0&1\\
0&1&0&0&0&1&0&0&1&0&1&0&0&1&1&1&0&1&0&0&0&0&0&0&0&0&1&1&0&0&1&0&0&0&0&1\\
0&0&0&0&0&0&0&0&1&1&0&0&1&0&0&0&0&1&0&1&0&0&0&1&0&0&0&1&1&0&1&1&1&1&0&0\\
0&0&1&0&0&1&0&1&1&1&0&0&0&0&0&1&1&1&0&0&0&0&0&0&0&0&1&1&1&0&1&0&0&1&0&0\\
0&0&0&0&0&0&0&0&1&1&1&0&1&0&0&1&0&0&0&0&1&0&0&1&0&1&0&0&1&0&1&0&0&0&1&1\\
0&0&0&1&0&1&0&1&0&1&1&1&0&0&1&0&1&0&0&0&0&0&0&0&0&0&1&1&1&1&0&0&0&1&1&0\\
0&0&0&0&0&0&0&0&1&1&1&1&0&0&0&1&1&0&0&0&0&1&0&1&0&1&1&0&0&0&0&0&1&1&0&0\\
0&0&0&0&1&1&0&0&1&1&0&1&0&1&0&0&1&1&0&0&0&0&0&0&0&1&0&0&0&0&0&1&0&1&1&1\\
0&0&0&0&0&0&0&1&0&0&0&0&0&1&0&1&1&1&0&0&0&0&1&1&0&1&1&1&0&1&0&0&0&1&0&0\\
0&0&0&0&0&0&1&1&1&1&1&1&0&0&0&0&0&0&0&0&0&0&0&0&0&0&0&0&0&0&1&1&1&1&1&1\\
0&0&0&0&0&0&0&0&0&0&0&0&1&1&1&1&1&1&0&0&0&0&0&0&1&1&1&1&1&1&1&1&1&1&1&1
\end{array}
\right).
\end{align}
\end{widetext}
As in the \B{the case of the $[[30,6,5]]$ code}, the corresponding \B{$[[36,12,5]]$} symplectic double code is constructed from \B{this matrix according to Eq.~(3) in the main text}.

By finding six vectors orthogonal to and linearly independent of both $H_{X}$ and $H_{Z}$, we obtain the explicit forms of $M_X$ and $M_Z$ as
\begin{align}
M_{X} = M_{Z} =
\begin{pmatrix} M & O_{6,18} \\ O_{6,18} & M \end{pmatrix},
\end{align}
where 
\begin{align}
M=\left(\begin{array}{cccccccccccccccccc}
1&0&0&0&0&0&0&1&1&1&1&0&0&0&1&1&0&0\\
0&1&0&0&0&0&0&0&1&1&1&1&0&0&0&1&1&0\\
0&0&1&0&0&0&1&0&0&1&1&1&0&0&0&0&1&1\\
0&0&0&1&0&0&1&1&0&0&1&1&1&0&0&0&0&1\\
0&0&0&0&1&0&1&1&1&0&0&1&1&1&0&0&0&0\\
0&0&0&0&0&1&1&1&1&1&0&0&0&1&1&0&0&0
\end{array}
\right).
\end{align}
consists of the above six vectors.
The logical Pauli operators are expressed more explicitly as follows:
\begin{align}
Z_{L1} &= Z_{1}Z_{8}Z_{9}Z_{10}Z_{11}Z_{15}Z_{16}, \\
X_{L1} &= X_{1}X_{8}X_{9}X_{10}X_{11}X_{15}X_{16}, \\
Z_{L2} &= Z_{2}Z_{9}Z_{10}Z_{11}Z_{12}Z_{16}Z_{17}, \\
X_{L2} &= X_{2}X_{9}X_{10}X_{11}X_{12}X_{16}X_{17}, \\
Z_{L3} &= Z_{3}Z_{7}Z_{10}Z_{11}Z_{12}Z_{17}Z_{18}, \\
X_{L3} &= X_{3}X_{7}X_{10}X_{11}X_{12}X_{17}X_{18},\\
Z_{L4} &= Z_{4}Z_{7}Z_{8}Z_{11}Z_{12}Z_{13}Z_{18}, \\
X_{L4} &= X_{4}X_{7}X_{8}X_{11}X_{12}X_{13}X_{18},\\
Z_{L5} &= Z_{5}Z_{7}Z_{8}Z_{9}Z_{12}Z_{13}Z_{14}, \\
X_{L5} &= X_{5}X_{7}X_{8}X_{9}X_{12}X_{13}X_{14},\\
Z_{L6} &= Z_{6}Z_{7}Z_{8}Z_{9}Z_{10}Z_{14}Z_{15}, \\
X_{L6} &= X_{6}X_{7}X_{8}X_{9}X_{10}X_{14}X_{15},\\
Z_{L7} &= Z_{19}Z_{26}Z_{27}Z_{28}Z_{29}Z_{33}Z_{34}, \\
X_{L7} &= X_{19}X_{26}X_{27}X_{28}X_{29}X_{33}X_{34}, \\
Z_{L8} &= Z_{20}Z_{27}Z_{28}Z_{29}Z_{30}Z_{34}Z_{35}, \\
X_{L8} &= X_{20}X_{27}X_{28}X_{29}X_{30}X_{34}X_{35},
\end{align}
\begin{align}
Z_{L9} &= Z_{21}Z_{25}Z_{28}Z_{29}Z_{30}Z_{35}Z_{36}, \\
X_{L9} &= X_{21}X_{25}X_{28}X_{29}X_{30}X_{35}X_{36},\\
Z_{L10} &= Z_{22}Z_{25}Z_{26}Z_{29}Z_{30}Z_{31}Z_{36}, \\
X_{L10} &= X_{22}X_{25}X_{26}X_{29}X_{30}X_{31}X_{36},\\
Z_{L11} &= Z_{23}Z_{25}Z_{26}Z_{27}Z_{30}Z_{31}Z_{32}, \\
X_{L11} &= X_{23}X_{25}X_{26}X_{27}X_{30}X_{31}X_{32},\\
Z_{L12} &= Z_{24}Z_{25}Z_{26}Z_{27}Z_{28}Z_{32}Z_{33}, \\
X_{L12} &= X_{24}X_{25}X_{26}X_{27}X_{28}X_{32}X_{33},
\end{align}
where $X_{Li}$ and $Z_{Li}$ are the Pauli operators for the $i$-th logical qubit, and $X_{j}$ and $Z_{j}$ are those for the $j$-th physical qubit.

\B{Figures~9--13 show the results for the $[[36,12,5]]$ code, each of which} 
corresponds directly to a figure in the main text.
Specifically, Fig.~9 corresponds to Fig.~2, Fig.~10 to Fig.~3, \B{Figs}.~11A and 11B to \B{Figs}.~5A and 5B, respectively, \B{Figs}.~12A and 12B to \B{Figs}.~7C and 7D, respectively, and Fig.~13 to Fig.~8. 
\B{These results are obtained by applying the proposed methods explained in the main text directly to the $[[36,12,5]]$ code. 
As shown in Fig.~4, the logical-CNOT error probabilities 
for the $[[36,12,5]]$ code are comparable to 
those for the $[[30,6,5]]$ code with a reasonable increase 
in qubit resources. 
This supports the generality of the proposed methods.} 


\begin{figure*}[htbp]
\centering
\includegraphics[width=0.65\columnwidth]{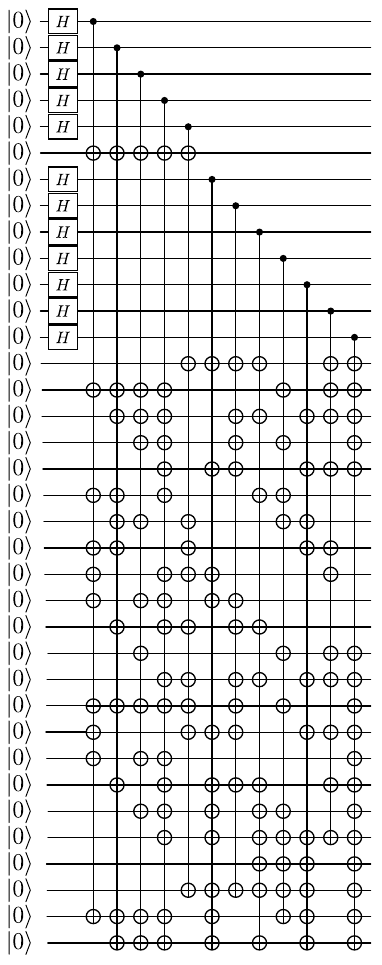}
\caption{\textbf{Zero-state encoder for the $[[36,12,5]]$ symplectic double code with $143$ CNOT gates.}}
\end{figure*}

\begin{figure*}[h]
\hspace*{-1cm}
\includegraphics[width=0.95\textwidth]{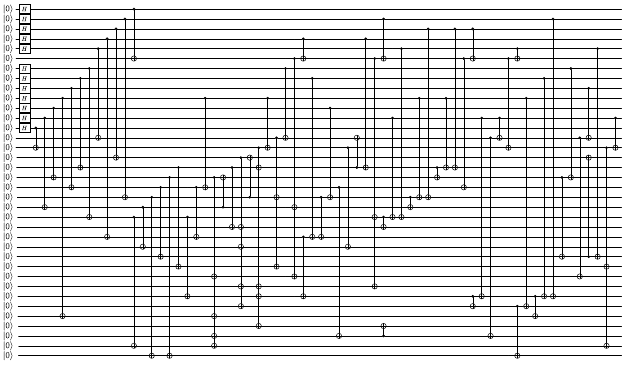}
\caption{\textbf{Zero-state encoder for the $[[36,12,5]]$ symplectic double code with $86$ CNOT gates.}}
\end{figure*}

\begin{figure*}[htbp]
\centering
\includegraphics[width=1.7\columnwidth]{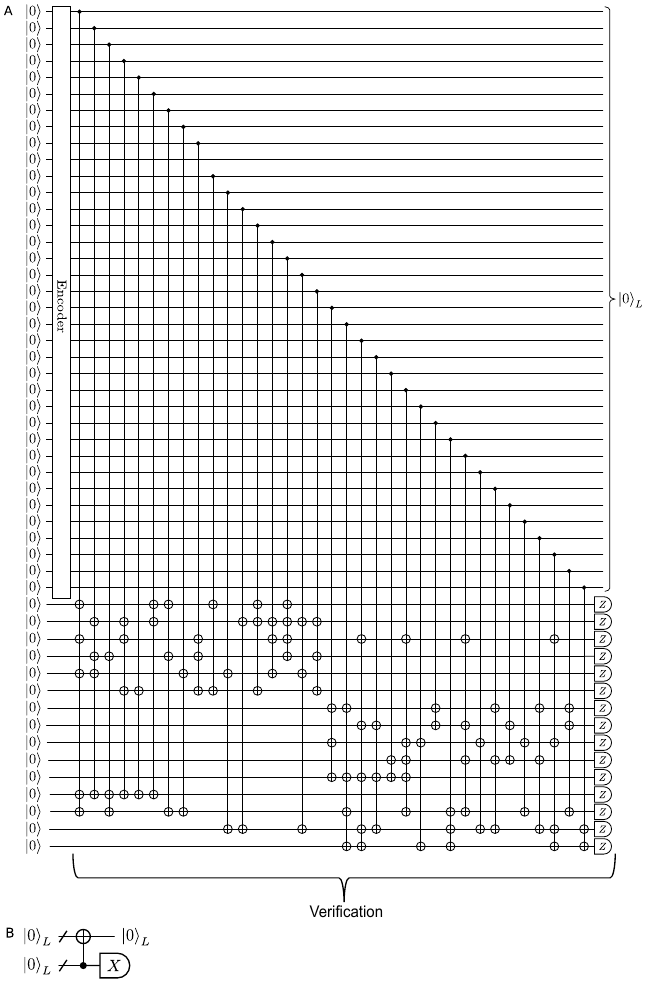}
\caption{\textbf{Fault-tolerant zero-state encoder for the $[[36,12,5]]$ symplectic double code.} 
(\textbf{A}) Zero-state encoder utilizing \Imp{Goto's verification method~\cite{17} to eliminate harmful $X$ errors}. 
The ``Encoder" box is the \Imp{quantum circuit} shown in FIG.~10. 
(\textbf{B}) \Imp{Verification procedure to eliminate harmful $Z$ errors. The two input states are prepared by the encoder in (A).}}
\end{figure*}

\begin{figure*}[htbp]
\centering
\includegraphics[width=1.3\columnwidth]{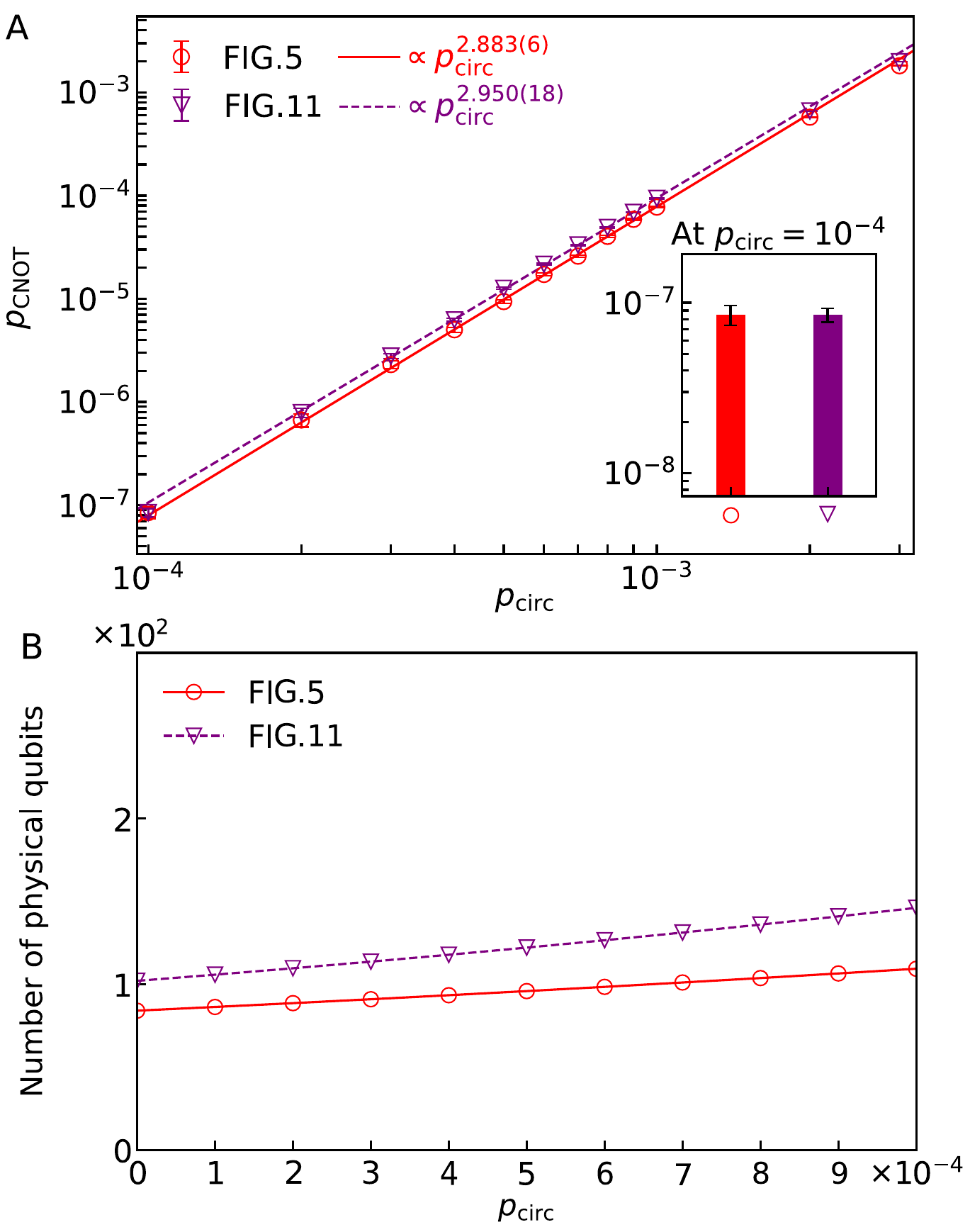}
\caption{\textbf{Evaluation of the encoders and logical CNOT gates 
\B{for the $[[36,12,5]]$ symplectic double code}.}
(\textbf{A})~Numerical results of the logical-CNOT error probability \B{$p_{\mathrm{CNOT}}$} as a function of the error rate, $p_{\mathrm{circ}}$, of the \B{circuit-level} noise model. 
\B{The circles and triangles show the results 
for the $[[30,6,5]]$ and $[[36,12,5]]$ symplectic double codes, respectively, 
where the proposed encoders in Figs.~5 and 11 are used, respectively.} 
\B{The scaling exponent for the $[[36,12,5]]$ code} was obtained by \B{using} the data over the \B{range of $p_{\mathrm{circ}}$ 
from $10^{-4}$ to $10^{-3}$}.
\B{The results for the $[[30,6,5]]$ code are the same as those in Fig.~7C in the main text.}
(\textbf{B})~\Blue{Average} total number of physical qubits required for preparing a logical \B{all-plus/all-zero} state 
\B{passing the verification stage with no detected errors}.}
\end{figure*}

\begin{figure*}[htbp]
\centering
\includegraphics[width=1.2\columnwidth]{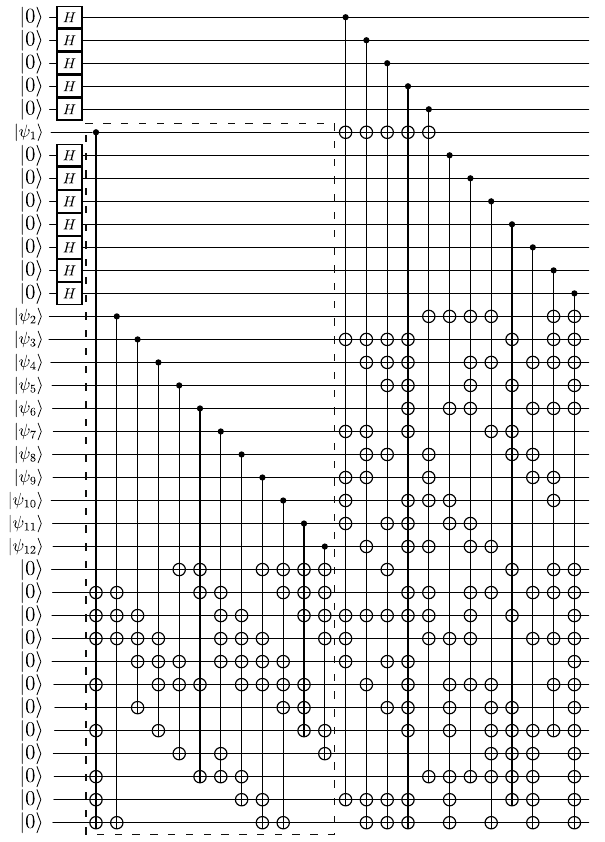}
\caption{\textbf{Arbitrary-state encoder for the $[[36,12,5]]$ symplectic double code.}}
\end{figure*}


\end{document}